\documentclass[pre,aps,superscriptaddress,showpacs]{revtex4}

\def \beq{\begin{equation}}         \def \eeq{\end{equation}}
\def \beqa{\begin{eqnarray}}        \def \eeqa{\end{eqnarray}}
\def \bea{\begin{array}}        \def \eea{\end{array}}

\usepackage{amsmath}
\usepackage{epsfig}
\usepackage[dvips]{color}

\begin{document}

%Title of paper
\title{Equivalence of two Bochkov-Kuzovlev equalities in quantum two-level systems}
\author{Fei Liu}
\email[Email address: ]{feiliu@buaa.edu.cn} \affiliation{School
of Physics and Nuclear Energy Engineering, Beihang University,
Beijing 100191, China}
\date{\today}

\begin{abstract}
{We present two kinds of Bochkov-Kuzovlev work equalities in a
two-level system that is described by a quantum Markovian master
equation. One is based on multiple time correlation functions and
the other is based on the quantum trajectory viewpoint. We show
that these two equalities are indeed equivalent. Importantly, this
equivalence provides us a way to calculate the probability density
function of the quantum work by solving the evolution equation for
its characteristic function. We use a numerical model to verify
these results. }
\end{abstract}
 \maketitle

{\noindent \it Introduction.} In the past decade, extending
classical fluctuation
relations~\cite{Evans93,Gallavotti95,Kurchan98,Lebowitz99,Bochkov77,JarzynskiPRL97,
Crooks99,HatanoSasa01,Maes03,SeifertPRL05,Speck05,Kawai07,EspositoPRL10}
into nonequilibrium quantum regime is attracting intensive
interest~\cite{Bochkov77,Kurchanquantum,Tasaki00,Yukawa00,PiechocinskaPRA00,MukamelPRL03,DeRoeck04,TalknerJPA07,TalknerPRE07,AndrieuxPrl08,Crooks08,Esposito09,
Deffer11,Campisi11,CampisiPTRS11,Hekking13,Horowitz12,Leggio13,Mazzola13,Dorner13}.
In the literature, the quantum
measurement~\cite{Kurchanquantum,Tasaki00,TalknerJPA07,TalknerPRE07,Campisi11}
and the quantum trajectory in Hilbert
space~\cite{Carmichael93,Breuer02,Wiseman10,Plenio98,Esposito09,DeRoeck04,Horowitz12,Hekking13,Leggio13}
are two widely used fundamental concepts. As an alternative to
those two concepts, very recently, Chetrite and
Mallick~\cite{Chetrite12}, and we~\cite{LiuFPRE12} showed that, in
the isolated Hamiltonian system and some quantum Markovian master
equations (QMME)~\cite{Breuer02,Carmichael02}, the quantum work
equalities~\cite{Bochkov77,JarzynskiPRL97} can be derived based on
the time-reversal and quantum Feynman-Kac formulas. In contrast
with the conventional work equalities written as statistical
average of exponential
functions~\cite{Esposito09,DeRoeck04,Horowitz12,Hekking13,Leggio13},
which we name them the {\it c-number} equalities, the newly
obtained equalities that are named the {\it q-number} equalities
are remarkable analogies with their classical
counterparts~\cite{LiuFPRE12,Chetrite12}. Even so, in the case of
the QMMEs, the exact relationship of the q-number and c-number
equalities, and whether the q-number equalities are useful besides
their forms have not been seriously considered. In this Rapid
communication, we use a driven quantum two-level system (TLS) with
dissipation to prove that, the q-number and the c-number
Bochkov-Kuzovlev equalities (BKE) are indeed equivalent for a
specific class of QMMEs. An important consequence of this
investigation is that we find an efficient way to calculate the
probability density function (pdf) of the quantum work for these
systems without doing the quantum jump simulation~\cite{Breuer02,Wiseman10}.\\

{{\noindent \it Driven quantum two-level system.} \label{section2}
The TLS has a free Hamiltonian $H_0$$=$$\hbar\omega\sigma_z/2$.
Initially, the system is in the thermal state
$\rho_0$$=$$\exp(-\beta H_0)/{\rm Tr}[\exp(-\beta H_0)]$ and
$\beta$ is the inverse temperature of the surrounding heat
reservoir. After time 0, a driving field is applied on the system
up to the final time $T$. During the whole process, we assume that
the evolution equation of the reduced density matrix of the system
$\rho(t)$ is
\begin{eqnarray}
\label{forward2LS}
\partial_{t}\rho(t)=L_{t}\rho(t)=-\frac{i}{\hbar}[H_0+H_1(t),\rho(t)]+D[\rho(t)],
\end{eqnarray}
where $H_1(t)$ is the interaction energy of the system and the
driving field and we do not need to specify its concrete
expression now. The time-independent term $D$ represents the
dissipation due to the interaction between the TLS and the heat
bath, which is
\begin{eqnarray}\label{dissipation}
D[\rho(t)]=\gamma_\downarrow(\sigma_-\rho\sigma_+- \frac{1}{2}
\{\rho,\sigma_+\sigma_-\})+
\gamma_\uparrow(\sigma_+\rho\sigma_--\frac{1}{2}\{\rho,\sigma_-\sigma_+\}),
\end{eqnarray}
where the two damping rates satisfy the detailed balance
condition~\cite{Breuer02},
\begin{eqnarray}
\label{detailedbalance}
\gamma_{\uparrow}=\gamma_{\downarrow}\exp(-\beta\hbar\omega).
\end{eqnarray}
This condition ensures that the system relaxes to the thermal
state $\rho_0$ if we switch off $H_1(t)$.
Equation~(\ref{forward2LS}) represents a class of QMMEs, in which
the coupling of the driving field to the system and the bath is
weak~\cite{Bloch56,Redfield57,Geva,Breuer97,Breuer04,Szczygielski,Slichter,Carmichael02}.
We must point out that, the model is distinct from those in
previous work~\cite{DeRoeck04,Esposito09,Chetrite12,Horowitz12}:
if one fixes the driving field at some value, the TLS may relax to
some steady state but generally not to the thermal state
$\propto\exp[-\beta(H_0+H_1)]$. The superoperator $D$ possesses an
important property~\cite{Spohn78}:
\begin{eqnarray}
\label{detailedbalance} D[A\rho_0]=D^\star[A]\rho_0,
\end{eqnarray}
where the dual of $D$ is
\begin{eqnarray}
\label{detailedbalanceoperatorform}
D^\star[A]=\gamma_\downarrow(\sigma_+A\sigma_-- \frac{1}{2}
\{A,\sigma_+\sigma_-\})+
\gamma_\uparrow(\sigma_-A\sigma_+-\frac{1}{2}\{A,\sigma_-\sigma_+\}).
\end{eqnarray}

{\noindent \it Q-number BKE.} \label{section3} Following the
spirit of establishing the classical work
equalities~\cite{LiuFJPA10,LiuFPRE09,Chetrite08}, we first
introduce the time-reversed process $\tilde\rho(s)$ of
Eq.~(\ref{forward2LS}). Its master equation is
\begin{eqnarray}
\label{reversedsystem}
\partial_s \tilde\rho(s)=\tilde{L}_s\tilde\rho(s)=-\frac{i}{\hbar}[H_0+\tilde H_1(s),\tilde\rho(s)]+D[\tilde\rho(s)],
\end{eqnarray}
where $H_0$ is time-reversible, $\tilde{H}_1(s)$$=$$\Theta
H_1(t')\Theta^{-1}$ with $t'$=$T$$-$$s$, and $\Theta$ is
time-reversal operator. We specifically set up the initial
condition of the reversed process to be $\rho_0$. The next step is
to obtain a solution for the operator $R(t',T)$ which is defined
as
\begin{eqnarray}
\label{Roperator} \tilde{\rho}(s)=\Theta R(t',T)\rho_0\Theta^{-1}.
\end{eqnarray}
$R(t',T)$ indicates the deviation of the perturbed system from the
equilibrium state $\rho_0$. Obviously, $R$$(T,T)$ is the identity
operator $I$. Substituting Eq.~(\ref{Roperator}) into
Eq.~(\ref{reversedsystem}) and using the
relationship~(\ref{detailedbalance}), we obtain an equation of
motion for $R(t',T)$ with respect to $t'$:
\begin{eqnarray}
\label{EOMR} \partial_{t'} R(t',T)&=&-L^\star_{t'} R(t',T) -
R(t',T)\hspace{0.1cm}\frac{i}{\hbar}[H_1(t'),\hspace{0.1cm}\rho_0]{\rho_0}^{-1}\\
&=&-L^\star_{t'} R(t',T)-{\cal W}_{t'} R(t',T),
\end{eqnarray}
where $L_{t'}^\star$ is the dual of $L_{t'}$~\cite{Breuer02}. We
also introduced the superoperator ${\cal W}_t$. Its action on an
operator is a multiplication from the right-hand side of the
operator. Using the celebrated Dyson series, Eq.~(\ref{EOMR}) has
the following formal solution~\cite{LiuFarxiv12,Chetrite12}:
\begin{eqnarray}
\label{DysonexpansionR}
R(t',T)=[G^\star(t',T)+\sum_{n=1}^{\infty}\int_{t'}^Tdt_1\cdots\int_{t_{n-1}}^T
dt_n\prod_{i=1}^n G^\star(t_{i-1},t_i){\cal W}_{t_i}
G^\star(t_n,T)] R(T,T),
\end{eqnarray}
where $G^\star(t_1,t_2)$$=$${\cal T}_{+}\exp [\int_{t_1}^{t_2}
d\tau L^*_{\tau}$] $(t_1$$<$$t_2)$ is the adjoint propagator of
the system, and ${\cal T}_{+}$ denotes the antichronological
time-ordering operator. Notice that
$G^\star(t_1,t_2)(I)$$=$$I$~\cite{Breuer02}.

Equation~(\ref{Roperator}) has a trivial property, {\it i.e.}, the
traces of its both sides being 1. Hence, substituting
Eq.~(\ref{DysonexpansionR}) and letting $t'$$=$$0$, we obtain the
q-number BKE:
\begin{eqnarray}
\label{BKEI} 1&=& {\rm Tr}[R(0,T)\rho_0]= 1+\int_0^Tdt_1{\rm
Tr}[(\frac{i}{\hbar}[H_1(t_1),\rho_0]\rho_0^{-1})G(t_1,0)(\rho_0)]\nonumber\\
\hspace{0.3cm}&&+\int_0^Tdt_1\int_{t_1}^Tdt_2{\rm
Tr}[(\frac{i}{\hbar}[H_1(t_2),\rho_0]\rho_0^{-1})G(t_2,t_1)(
\frac{i}{\hbar}[H_1(t_1),\rho_0]\rho_0^{-1}G(t_1,0)(\rho_0))]+\cdots\nonumber
\\&=&1+ \int_0^Tdt_1\langle
(\frac{i}{\hbar}[H_1(t_1),\rho_0]\rho_0^{-1}) \rangle +
\int_0^Tdt_1\int_{t_1}^T dt_2\langle (
\frac{i}{\hbar}[H_1(t_2),\rho_0]\rho_0^{-1})(
\frac{i}{\hbar}[H_1(t_1),\rho_0]\rho_0^{-1}) \rangle+\cdots\nonumber\\
&=&\langle\hspace{0.1cm} {\cal T}_+\exp(\int_0^T d\tau
\frac{i}{\hbar}[H_1(\tau),\rho_0]\rho_0^{-1})\hspace{0.1cm}\rangle,
\end{eqnarray}
where $G(t_2,t_1)$$=$${\cal T}_{-}\exp[ \int_{t_1}^{t_2} d\tau
L_{\tau}]$ is the system's propagator from time $t_1$ to $t_2$,
and ${\cal T}_{-}$ denotes the chronological time-ordering
operator. We have used the property ${\rm
Tr}[G^\star(t_1,t_2)(A)B]={\rm Tr}[A
G(t_2,t_1)(B)]$~\cite{Breuer02}. We denote the form of the
right-hand side of Eq.~(\ref{BKEI}) the quantum Feynman-Kac
formula~\cite{Chetrite12,LiuFPRE12}. We must remind the reader
that the notation $\langle$$\cdots$$\rangle$ or the ``average"
above is a shorthand of the multi-time correlation functions of
the operators~\cite{Breuer02}. In the absence of the dissipation
term $D$, Eq~(\ref{BKEI}) reduces into the quantum BKE for the
isolated Hamiltonian system~\cite{CampisiPTRS11,LiuFPRE12}.
Moreover, if we interpret $-i[\cdots]/\hbar$ as Poisson bracket,
the density matrixes as distribution functions, and the
propagators under classical meaning, Eq.~(\ref{BKEI}) then becomes
the classical BK
equality~\cite{Bochkov77,Jarzynskicmp,LiuFJPA10,HorowitzJSM07}. It
is worthy emphasizing that the concrete formulas of $H_0$ and $D$
are not involved in the above discussion.\\

{{\noindent \it C-number BKE.} \label{section4} On the basis of
the theory of quantum jump trajectory, we may obtain an
alternative quantum BK equality~\cite{comment1}. Since the basic
idea and techniques have been given
previously~\cite{Esposito09,Hekking13,DeRoeck04,Crooks08}, here we
only present the essential ingredients. According to the
theory~\cite{Carmichael93,Plenio98,Wiseman10,Breuer02}, the state
vector $\psi(t)$ of the reduced TLS system evolves in its Hilbert
space by deterministic continuous evolution and stochastic jumps
alternatively. The deterministic equation is
\begin{eqnarray}
\label{determinedwavefunction}
\partial_t\psi(t)&=&-\frac{i}{\hbar}\hat{H}(t)\psi(t)=-\frac{i}{\hbar}[H_0+H_1(t) -
\frac{i}{2}\hbar(\gamma_\downarrow\sigma_{+}\sigma_{-}+\gamma_\uparrow\sigma_{-}\sigma_{+})]\psi(t).
\end{eqnarray}
The equation has a formal solution,
$\psi(t)$$=$$U(t,t_1)\psi(t_1)/\|\psi(t_1)\|^2$, where the
non-unitary time evolution operator $U(t,t_1)$ is ${\cal
T}_{-}\exp[-\frac{i}{\hbar}\int_{t_1}^td\tau \hat{H}(\tau)].$
Occasionally, this evolution is interrupted by a stochastic jump
to one of the states: $\sigma_{+}\psi(t)/\|\sigma_{+}\psi(t)\|^2$
and $\sigma_{-}\psi(t)/\|\sigma_{-}\psi(t)\|^2$. For the TLS these
are the excited state $|e\rangle$ and ground state $|g\rangle$,
respectively. In the quantum optics, these jumps appear an
absorbtion or emission of a
photon~\cite{Carmichael93,Wiseman10,Breuer02,Plenio98}. Hence, the
corresponding energy could be physically interpreted as heat
absorbed or released by the system from or to the heat
bath~\cite{Hekking13,Horowitz12,DeRoeck04,Crooks08,Esposito09}. By
measuring the energy of the TLS at the beginning time
($\epsilon_n$) and ending time ($\epsilon_m$) while recording the
number $N_+$ ($N_-$) of the jumps to $|e\rangle$ ($|g\rangle$)
along a quantum trajectory, we define the work done by the driving
field on the TLS as
\begin{eqnarray}
\label{workdef} W=\epsilon_n-\epsilon_m-\omega\hbar
\int_0^TdN_++\omega\hbar\int_0^T dN_-,
\end{eqnarray}
where $dN_+$ and $dN_-$ are the increments of these two types of
jumps. We remind the reader that the first two terms are the
energy eigenvalues of the free Hamiltonian $H_0$ instead of the
total Hamiltonian. With the above notations, we give the c-number
BKE for the quantum work~(\ref{workdef}):
\begin{eqnarray}
\label{BKEII} 1&=&\sum_{m,n}p_m(0)\sum_{N=0}^\infty
\int_0^T\cdots\int_{t_{N-1}}^T\prod_{1}^Ndt_i
\sum_{\gamma_1}\cdots\sum_{\gamma_N}[\hspace{0.1cm}\prod_{1}^N
\gamma_i \left |\langle n|{\cal L}_N |m\rangle \right
|^2\hspace{0.1cm}] \hspace{0.2cm} e^{-\beta W }=E[ e^{-\beta W}],
\end{eqnarray}
where $p_m(0)$ $=$ $\exp(-\beta\epsilon_m)/{\rm Tr}[\exp(-\beta
H_0)]$ is the initial probability of the TLS at the eigenstate
with the energy $\varepsilon_m$, the whole term inside the square
brackets of the first equation is the probability density of
observing a quantum trajectory that starts from the eigenstate
$|m\rangle$, occurs jump at time $t_i$ with type $P_i$ that equals
$\sigma_+$ or $\sigma_-$ with the jump rate
$\gamma_i$=$\gamma_\uparrow$ or $\gamma_\downarrow$
($i$$=$$1$,$\cdots$,$N$), and ends in the eigenstate $|n\rangle$
at the final time $T$, and
\begin{eqnarray}
\label{timeevolutionoperator} {\cal L}_N=U(T,t_{N})P_{N}\cdots
U(t_2,t_1)P_1 U(t_1,0)
\end{eqnarray}
is the time evolution operator of the whole
trajectory~\cite{Breuer02}. We specifically use the notation
$E[\cdots]$ to denote the average in the c-number equality.
Proof of the equality will be seen shortly.\\

{{\noindent \it Equivalence of the two BKEs.} \label{section5}
Although we name Eq.~(\ref{BKEI}) the BKE, its physical relevance
is not obvious. We do not see from the abstract equality what the
work is and whether the second law of thermodynamics is implied.
It is quite different from the c-number BKE~(\ref{BKEII}). At
first glance, these two equalities appear so distinct. However, we
will show that it is only superficial. Before the summation over
$m$, Eq.~(\ref{BKEII}) can be rewritten as
\begin{eqnarray}
\label{proof} &&p_m(0)\sum_{n}\sum_{N=0}^\infty
\int_0^T\cdots\int_{t_{N-1}}^T\prod_{1}^Ndt_i
\sum_{\gamma_1}\cdots\sum_{\gamma_N}[\hspace{0.1cm}\prod_{1}^N
\tilde{\gamma_i} \left |\langle m|{\hspace{0.1cm}}{\cal L}_N^\dag
|n\rangle \right |^2\hspace{0.1cm}] \hspace{0.2cm} e^{-\beta W
}e^{-\beta(\hbar\omega N_+-\hbar\omega N_-)}\nonumber\\
&=&\langle m| \hspace{0.05cm}\Theta^{-1}\sum_{n}
p_n(0)\sum_{N=0}^\infty
\int_0^T\cdots\int_{s_{N-1}}^T\prod_{1}^Nds_i
\sum_{\tilde{\gamma}_1}\cdots\sum_{\tilde{\gamma}_N}\hspace{0.1cm}\prod_{1}^N
{\tilde{\gamma}}_i \hspace{0.1cm}\tilde{{\cal L}}_N
\hspace{0.1cm}\Theta |n\rangle\langle n|
\Theta^{-1}\hspace{0.1cm}\tilde{{\cal L}}_N^\dag  \hspace{0.1cm}
\Theta|m\rangle\nonumber\\
&=&\langle m| \Theta^{-1}
\tilde{\rho}(T)\hspace{0.1cm}\Theta|m\rangle,
\end{eqnarray}
where $\tilde{\gamma}_\downarrow$$=$$\gamma_\uparrow$,
$\tilde{\gamma}_\uparrow$$=$$\gamma_\downarrow$,
$s_i$$+$$t_{N+1-i}$$=$$T$,
\begin{eqnarray}
\tilde{{\cal L}}_N=\tilde{U}(T,s_{N})P_{1}^\dag\cdots \tilde
{U}(s_2,s_1)P_N^\dag \tilde{U}(s_1,0)
\end{eqnarray}
is the time evolution operator of the reversed quantum trajectory,
and $\tilde{U}(s,s_1)$ is analogous to the previous $U(t,t_1)$
except that the Hamiltonian therein is replaced by
$H_0$$+$$\tilde{H}_1(s)$. The last exponential term in the first
line of Eq.~(\ref{proof}) is the consequence of the detailed
balance condition~(\ref{detailedbalance}), and the final equation
is due to the well-established relationship between the density
matrix and the quantum trajectory~\cite{Breuer02,Wiseman10}.
Comparing Eq.~(\ref{Roperator}) with Eq.~(\ref{proof}), we
immediately see that, the whole expression after $p_m(0)$ is just
$\langle m|R(0,T)|m\rangle$ on the left hand side of the latter
equation. Therefore, we prove that the c-number and q-number BKEs
are exactly equivalent.

An alternative proof of this equivalence that does not depend on
the time-reversal explanation is to do series expansions for these
two BKEs in terms of $\beta$. We then compare their respective
coefficients of the different orders of $\beta$. For the c-number
BKE, the expansion is simply
\begin{eqnarray}
1=1-E[W]\beta  +\frac{1}{2}E[W^2]\beta^2 \cdots.
\end{eqnarray}
Using the facts that $E[dN_+]$$=$$\gamma_\uparrow$${\rm
Tr}[\sigma_-\sigma_+\rho(t)]dt$ and
$E[dN_-]$$=$$\gamma_\downarrow$${\rm
Tr}[\sigma_+\sigma_-\rho(t)]dt$~\cite{Breuer02,Wiseman10}, where
$t$ is the time of non-vanishing $dN_{\pm}$, we rewrite the first
moment of the work~(\ref{workdef}) as (see the Supplemental
Material)
\begin{eqnarray}
\label{firstmoment} E[W]&=&\int_0^T dt_1
\frac{d}{dt_1}\hspace{0.1cm}{\rm Tr}[H_0\rho(t_1)]-\int_0^Tdt_1
{\rm Tr}[D^\star[H_0]\rho(t_1)]=\int_0^Tdt_1\langle
\frac{i}{\hbar}[H_1(t_1),H_0]\rangle.
\end{eqnarray}
Because the left hand side is the average work and the first
integration in the first equation represents a change of average
energy of the TLS during the whole process, we may interpret the
second integration in the same equation as the absorbed average
heat from the heat bath. Hence, Eq.~(\ref{firstmoment}) is just
the {\it first law} of thermodynamics. Using the Jensen's
inequality, we surely have the {\it second law} of thermodynamics,
$E[W]\ge0$. A more complex case is the second moment. Using the
three crucial identities below~\cite{comment2},
\begin{eqnarray}
\label{correlation1}
E[dN_+dN'_+]&=&\{\gamma_\uparrow^2\hspace{0.1cm}{\rm
Tr}[\sigma_-\sigma_+G(t,t')(\sigma_+\rho(t')\sigma_-)]+\delta(t-t')\gamma_\uparrow{\rm
Tr}[\sigma_-\sigma_+\rho(t)]\}dtdt',\hspace{0.2cm} (t\ge t'),\\
\label{correlation2}
E[dN_-dN'_-]&=&\{\gamma_\downarrow^2\hspace{0.1cm}{\rm
Tr}[\sigma_+\sigma_-G(t,t')(\sigma_-\rho(t')\sigma_+)]+\delta(t-t')\gamma_\downarrow{\rm
Tr}[\sigma_+\sigma_-\rho(t)]\}dtdt',\hspace{0.2cm} (t\ge t'),\\
\label{correlation3}
E[dN_+dN'_-]&=&\{\gamma_\downarrow\gamma_\uparrow{\rm
Tr}[\sigma_-\sigma_+G(t,t')(\sigma_-\rho(t')\sigma_+)]\theta(t-t')+\gamma_\downarrow\gamma_\uparrow{\rm
Tr}[\sigma_+\sigma_-G(t',t)(\sigma_+\rho(t)\sigma_-)]\theta(t'-t)\}dtdt',
\end{eqnarray}
where $t$ ($t'$) is the time of non-vanishing $dN_{\pm}$
($dN'_{\pm}$), and doing a careful calculation, we obtain
\begin{eqnarray}
\label{secondmoment} \frac{1}{2}E[
W^2]=\int_0^Tdt_1\int_{t_1}^Tdt_2\langle
(\frac{i}{\hbar}[H_1(t_2),H_0])(
\frac{i}{\hbar}[H_1(t_1),H_0])\rangle-
\frac{1}{2}\int_0^Tdt_1\langle\hspace{0.1cm}[\hspace{0.1cm}\frac{i}{\hbar}[H(t_1),H_0],H_0]\rangle.
\end{eqnarray}
When we expand the q-number BKE~(\ref{BKEI}) accordingly, we find
that the coefficients of $\beta$ and $\beta^2$ are indeed the
right hand sides of Eqs.~(\ref{firstmoment})
and~(\ref{secondmoment}). Higher orders of $\beta$ can be checked
analogously. But the calculation becomes very long and tedious dramatically.\\

{{\noindent \it Characteristic function of the work.} The
preceding argument about the equivalence of Eq.~(\ref{BKEI}) and
Eq.~(\ref{BKEII}) is useful. First, we may apply
Eqs.~(\ref{firstmoment}) and~(\ref{secondmoment}) to calculate the
first two moments of the work by analytically or numerically
solving the master equations rather than by doing the quantum jump
simulation. Compared with the latter, the former is exact and
involves no sampling errors. As an illustration, we recalculate
these moments for the TLS model in Ref.~\cite{Hekking13}, where
$H_1(t)$$=$$\lambda_0\sin(\omega t)(\sigma_++\sigma_-)$; see
Fig.~(\ref{figure1}). The simulation data are also listed for a
comparison. Second, the equivalence provides us an interesting
method to calculate the pdf of the quantum work~(\ref{workdef}).
Letting the characteristic function~\cite{Campisi11} of the pdf be
$\Phi(u)$, where $u$ is the real number, we easily see that
\begin{eqnarray}
\label{characterfun} \Phi(u)=E[\hspace{0.05cm}e^{iuW}]={\rm
Tr}[K(0,T;u)\rho_0],
\end{eqnarray}
if the newly introduced operator $K(t',T;u)$ satisfies the
evolution equation given by
\begin{eqnarray} \label{evolutioneqcharactersiticfuncBKE}
\left\{%
\begin{array}{ll}
\partial_{t'}K(t',T;u)=-L^\star_{t'} K(t',T;u) -
K(t',T;u)\hspace{0.1cm}\frac{i}{\hbar}[H_1(t'),\hspace{0.1cm}e^{iuH_0}]e^{-iuH_0} ,& \\
K(T,T;u)=I.
\end{array}
\right.
\end{eqnarray}
By numerically solving the above equation and performing an
inverse Fourier transform of $\Phi(u)$, the pdf of the work is
then obtained. The inset of Fig.~(\ref{figure1}) is an example. We
see that our calculation agrees with the simulation data~\cite{Hekking13} excellently.\\
\begin{figure}
\includegraphics[width=0.8\columnwidth]{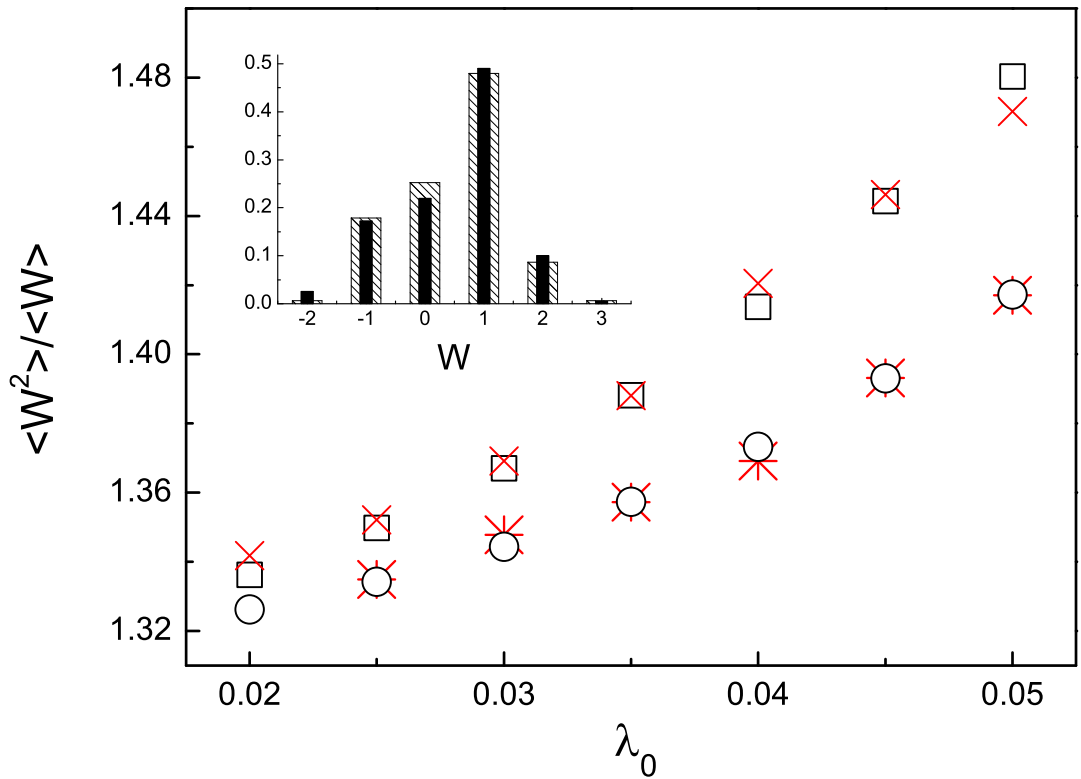}
\caption{The ratio of the second and first moments of the quantum
work (in unit $\hbar\omega$) with respect to different
perturbation strength $\lambda_0$ (in unit $\hbar\omega$) for the
TLS model in Ref.~\cite{Hekking13}, where $T\omega/2\pi$$=$$10$,
$\beta\hbar\omega$$=$$2.0$. The crosses
($\gamma_\downarrow$$=$$0.02\omega$) and stars
($\gamma_\downarrow$$=$$0.01\omega$) are the data of the quantum
jump simulation~\cite{Hekking13}, while the open squares and
circles are the numerical results of Eqs.~(\ref{firstmoment})
and~(\ref{secondmoment}). The inset shows the pdf of the quantum
work. The dashed bars are from the simulation of
Ref.~\cite{Hekking13}, and the solid black bars are obtained by
our characteristic function method, where $\beta\hbar
\omega$$=$$1.0$, $\gamma_\downarrow$$=$$0.05\omega$,
$\lambda_0$$=$$0.05\hbar\omega$. }\label{figure1}
\end{figure}

{{\noindent \it  Conclusion.}\label{section6} In this work, we
present two kinds of BKEs in the quantum TLS driven by the field
and we prove their equivalence. Moreover, an efficient way of
calculating the characteristic function of the quantum work is
revealed. So far, our discussions are limited to these specific
QMMEs where the driven field is so weak that their dissipations
can be treated as time-independent. Extending the current idea
into the more general cases, {\it e.g.}, the master equations with
time-dependent dissipations shall be very intriguing. We expect
that some of them would be related to the
quantum Jarzynski equality. This study is underway.\\

{\noindent We appreciate Prof. Hekking for permitting us to use
their simulation data in Ref.~\cite{Hekking13}. We also thank
Prof. Jarzynski, Dr. Deffner, and Zhiyue Lu for their useful
remarks on the work. This work was supported by the National
Science Foundation of China under Grant No. 11174025.}

\newpage
%\title{Supplemental Material for ``Equivalence of two Bochkov-Kuzovlev equalities in quantum two-level systems"}
%\maketitle

\section{Derivations of Eqs. (19) and (23)} For Eq.~(19), the
situation is simple:
\begin{eqnarray}
\label{firstmomentdef}
E[W]&=&E[\epsilon_n]-E[\epsilon_m]+\hbar\omega E[
N_+]-\hbar\omega E[N_-]\nonumber\\
&=&{\rm Tr}[H_0\rho(T)]-{\rm Tr}[H_0\rho(0)] +\int_0^T{\rm
Tr}[\hbar\omega(\gamma_\downarrow \sigma_+\sigma_- -
\gamma_\uparrow\sigma_-\sigma_+)\rho(t_1)]dt_1 \nonumber\\
&=&\int_0^Tdt_1 \frac{d}{dt_1}{\rm Tr}[H_0\rho(t_1)]-
\int_0^Tdt_1{\rm
Tr}[D^\star[H_0]\rho(t_1)]\nonumber\\&=&\int_0^Tdt_1 {\rm Tr}[
\frac{i}{\hbar}[H_1(t_1),H_0]\rho(t_1)].
\end{eqnarray}
For Eq.~(23), however, the proof becomes very tricky. First we
write down the original definition of the second moment of the
quantum work~(13),
\begin{eqnarray}
\label{2ndmomentdef}
E[W^2]&=&E[\epsilon_n^2+\epsilon_m^2-2\epsilon_n\epsilon_m]+2\hbar\omega
E[\epsilon_nN_+-\epsilon_nN_--\epsilon_mN_++\epsilon_mN_-]\nonumber\\&&+
(\hbar\omega)^2E[N_+^2- 2N_+ N_-+N_-^2].
\end{eqnarray}
The first two averages can be rewritten using the density matrix
$\rho(t)$ and the propagator $G(t_2,t_1)$ as
\begin{eqnarray}
\label{2ndmomentA} {\rm Tr}[H_0^2\rho(T)]+{\rm
Tr}[H_0^2\rho(0)]-2{\rm Tr}[H_0G(T,0)H_0\rho(0)]
\end{eqnarray}
and
\begin{eqnarray}
\label{2ndmomentB} &&\int_0^T dt_1{\rm Tr}[\gamma_\downarrow
H_0G(T,t_1)\sigma_-\rho(t_1)\sigma_+]-\int_0^Tdt_1{\rm
Tr}[\gamma_{\uparrow}
H_0G(T,t_1)\sigma_+\rho(t_1)\sigma_-]\nonumber\\
&&-\int_0^Tdt_1{\rm Tr}[\gamma_{\downarrow}
\sigma_+\sigma_-G(t_1,0)H_0\rho(0)]+\int_0^Tdt_1{\rm
Tr}[\gamma_{\uparrow} \sigma_-\sigma_+G(t_1,0)H_0\rho(0)],
\end{eqnarray}
respectively. For the last average in Eq.~(\ref{2ndmomentdef}), we
have to resort to Eqs.~(20)-(22) and obtain
\begin{eqnarray}
\label{2ndmomentC} &&2\int_0^Tdt_1\int_{t_1}^Tdt_2{\rm
Tr}[\gamma_{\downarrow}^2\sigma_+\sigma_-G(t_2,t_1)\sigma_-\rho(t_1)\sigma_+]+\int_0^Tdt_1{\rm
Tr}[\gamma_{\downarrow}\sigma_+\sigma_-\rho(t_1)]\nonumber\\
&&-2\int_0^Tdt_1\int_{t_1}^Tdt_2 {\rm
Tr}[\gamma_{\uparrow}\gamma_{\downarrow}\sigma_+\sigma_-G(t_2,t_1)\sigma_+\rho(t_1)\sigma_-]-2\int_0^Tdt_1\int_{t_1}^Tdt_2
{\rm
Tr}[\gamma_{\downarrow}\gamma_{\uparrow}\sigma_-\sigma_+G(t_2,t_1)\sigma_-\rho(t_1)\sigma_+]\nonumber\\
&&+2\int_0^Tdt_1\int_{t_1}^Tdt_2{\rm
Tr}[\gamma_{\uparrow}^2\sigma_-\sigma_+G(t_2,t_1)\sigma_+\rho(t_1)\sigma_-]+\int_0^Tdt_1{\rm
Tr}[\gamma_{\uparrow}\sigma_-\sigma_+\rho(t_1)].
\end{eqnarray}
Substituting Eqs.~(\ref{2ndmomentA})-(\ref{2ndmomentC}) into
Eq.~(\ref{2ndmomentdef}) and doing a rearrangement, we have
\begin{eqnarray}
\label{part1}
E[W^2]=&&2(\hbar\omega)^2\int_0^Tdt_1\int_{t_1}^Tdt_2{\rm
Tr}[(\gamma_{\downarrow}\sigma_+\sigma_--\gamma_{\uparrow}\sigma_-\sigma_+)
G(t_2,t_1)(\gamma_{\downarrow}\sigma_-\rho(t_1)\sigma_+
-\gamma_{\uparrow}\sigma_+\rho(t_1)\sigma_-)]\nonumber\\
&&+2\hbar\omega\int_0^T dt_1{\rm Tr}[
H_0G(T,t_1)(\gamma_\downarrow\sigma_-\rho(t_1)\sigma_+-
\gamma_{\uparrow}\sigma_+\rho(t_1)\sigma_-)]\nonumber \\
&&-2\hbar\omega\int_0^Tdt_1{\rm Tr}[(\gamma_{\downarrow}
\sigma_+\sigma_- - \gamma_{\uparrow}
\sigma_-\sigma_+)G(t_1,0)H_0\rho(0)]\nonumber\\
 &&+{\rm Tr}[H_0^2\rho(T)]+{\rm
Tr}[H_0^2\rho(0)]-2{\rm
Tr}[H_0G(T,0)H_0\rho(0)]\nonumber\\
 &&+(\hbar\omega)^2\int_0^T dt_1{\rm
Tr}[\gamma_{\downarrow}\sigma_+\sigma_-\rho(t_1)]+(\hbar\omega)^2\int_0^Tdt_1{\rm
Tr}[\gamma_{\uparrow}\sigma_-\sigma_+\rho(t_1)].
\end{eqnarray}
At this step, we do not see that Eq.~(\ref{part1}) essentially
equals to the right hand side of Eq.~(23). In order to go head, We
need to introduce two additional equations:
\begin{eqnarray}L_t[H_0\rho]&=&H_0L_t[\rho]-\frac{i}{\hbar}[H_1(t),H_0]\rho+\hbar\omega (\gamma_{\downarrow}\sigma_-\rho\sigma_+-
\gamma_{\uparrow}\sigma_+\rho\sigma_-),\\
L_t^\star[H_0H_0]&=&\frac{i}{\hbar}[H_1(t),H_0]H_0+\frac{i}{\hbar}H_0[H_1(t),H_0]+2D^\star[H_0]H_0+
(\hbar\omega)^2\gamma_{\downarrow}\sigma_+\sigma_-+
(\hbar\omega)^2\gamma_{\uparrow}\sigma_-\sigma_+.
\end{eqnarray}
Using the expression of $D^\star[H_0]$ in
Eq.~(\ref{firstmomentdef}), the definition of the adjoint
propagator $G^\star(t_1,t_2)$, the above two equations, and
carrying out further calculations we obtain
\begin{eqnarray}
\frac{1}{2}E[W^2]=&&-\int_0^Tdt_1\int_{t_1}^T dt_2 {\rm
Tr}[D^\star[H_0]G(t_2,t_1)\frac{i}{\hbar}[H_1(t_1),H_0]\rho(t_1)]\nonumber\\
&&+\int_0^Tdt_1{\rm
Tr}[H_0G(T,t_1)\frac{i}{\hbar}[H_1(t_1),H_0]\rho(t_1)]\nonumber\\
&&-\frac{1}{2}\int_0^Tdt_1{\rm
Tr}[\frac{i}{\hbar}[H_1(t_1),H_0]H_0\rho(t_1)]-\frac{1}{2}\int_0^Tdt_1{\rm
Tr}[H_0\frac{i}{\hbar}[H_1(t_1),H_0]\rho(t_1)].
\end{eqnarray}
Using the property of $G^\star(t_1,t_2)$,
\begin{eqnarray}
\partial_{t_2}[G^\star(t_1,t_2)H_0]=G^\star(t_1,t_2)\frac{i}{\hbar}[H_1(t_1),H_0]+G^\star(t_1,t_2)D^\star[H_0],
\end{eqnarray}
we finally arrive at the right hand side of Eq.~(23). Noting that
$G^\star(t_1,t_2)$ is a superoperator that acts on the operator on
its right hand side~\cite{Breuer02}.

\section{Calculating $K(t',T;u)$ for the TSL model} For the
simple resonant TSL model in Ref.~\cite{Hekking13}, we may write
the operator $K(t',T;u)$ in the Pauli matrixes as
\begin{eqnarray}
K(t',T;u)=\frac{1}{2} [k_0(t') I +k_x(t')\sigma_x + +
k_y(t')\sigma_y+ k_z(t')\sigma_z].
\end{eqnarray}
Substituting it into Eq.~(25) and doing a simple derivation, we
obtain
\begin{eqnarray}
\dot{k_0}
&=&\frac{i}{2}e^{-iu}(e^{iu}-1)^2\lambda(t')k_x-\frac{1}{2}e^{-iu}(e^{2iu}-1)\lambda(t')k_y+(\gamma_{\downarrow}-
\gamma_{\uparrow})k_z,\\
\dot{k_x}&=&\frac{i}{2}e^{-iu}(e^{iu}-1)^2\lambda(t')k_0+\frac{1}{2}(\gamma_{\downarrow}+\gamma_{\uparrow})k_x-\omega
k_y
+\frac{i}{2}e^{-iu}(e^{2iu}-1)\lambda(t')k_z,\\
\dot{k_y}&=&\frac{1}{2}e^{-iu}(1-e^{2iu}) \lambda(t')k_0+\omega
k_x+\frac{1}{2}(\gamma_{\downarrow}+\gamma_{\uparrow})k_y-\frac{1}{2}e^{-iu}(1+e^{iu})^2\lambda(t')k_z,\\
\dot{k_z}&=&\frac{i}{2}e^{-iu}(1-e^{2iu})\lambda(t')k_x+\frac{1}{2}e^{-iu}(1+e^{iu})^2\lambda(t')k_y+(\gamma_{\downarrow}+
\gamma_{\uparrow})k_z,
\end{eqnarray}
where the dots denote the time derivative $d/d{t'}$,
$\lambda(t')$$=$$\lambda_0\omega\sin(\omega t')$, and the terminal
conditions are $k_0(T)$$=$$2$,
$k_x(T)$$=$$k_y(T)$$=$$k_z(T)$$=$$0$, respectively. The reader is
reminded that all parameters are dimensionless. We clearly see
that the operator $K$ is periodic with respect to $u$, {\it i.e.}
$K(t',T;u+2n\pi)$$=$$K(t',T;u)$ for arbitrary integer $n$. This
feature ensures that the pdf of the quantum work is discrete after
we perform the inverse Fourier transform for $\Phi(u)$. These
differential equations can be easily solved numerically as a
terminal problem, {\it e.g.}, by using the Mathematica.

\section{General QMMEs having structure of Eq.~(1)} We have
mentioned that Eq.~(1) is a simplest example of the specific type
of QMMEs. The dissipation parts of these QMMEs have the following
common structure~\cite{Breuer02}
\begin{eqnarray}
\label{generalQMMEs} D[\rho]=\sum_j \gamma^j_\downarrow(A^j_-\rho
A^j_+- \frac{1}{2} \{\rho,A^j_+A^j_-\})+
\gamma^j_\uparrow(A^j_+\rho A^j_--\frac{1}{2}\{\rho,A^j_-A^j_+\}),
\end{eqnarray}
where the Lindblad operators are the eigenoperators of the free
Hamiltonian $H_0$, {\it i.e.},
$[H_0,A^j_{\pm}]$$=$$\pm\hbar\omega_j A^j_{\pm}$, and the damping
rates are assumed to satisfy
$\gamma^j_{\uparrow}$$=$$\gamma^j_{\downarrow}\exp(-\beta\hbar\omega_j)
$. Except for the additional summation over all possible coupling
channels $j$ of the system with the heat bath, we do not see that
there are fundamental differences between the generalized and the
simplest QMMEs. Therefore, all general results in the main text
could be simply extended into the general situation by changing
$\sigma_{\pm}$$\rightarrow$$A^i_{\pm}$,
$N_{\pm}$$\rightarrow$$N^i_{\pm}$,
$\omega$$\rightarrow$$\omega_i$,
$\gamma_{\downarrow\uparrow}$$\rightarrow$$\gamma^i_{\downarrow\uparrow}$,
and doing appropriate summation over the various channels $j$,
{\it e.g.}, the quantum work for the QMMEs with
Eq.~(\ref{generalQMMEs}) is
\begin{eqnarray}
\label{workdefgeneral} W=\epsilon_n-\epsilon_m+
\sum_j\hbar\omega_jN^j_+-\sum_j\hbar\omega_j N^j_-.
\end{eqnarray}
Noting that the three Eqs.~(20)-(22) are not zero only for the
same channels.

\iffalse

\fi

\begin{thebibliography}{99}
\bibitem{comment1} Hekking and Pekola should first present the
c-number BKE~\cite{Hekking13}. However, they claimed that they
verified the quantum Jarzynski equality. Compared with the method
developed in this work, Their method is relatively complex.
\bibitem{comment2} These equations are the generalizatoins of
Eq. (4.50) in the book of Wiseman and Milburn~\cite{Wiseman10}.
\end{thebibliography}
\end{document}